\documentclass[numberedappendix]{emulateapj} 
\usepackage{natbib}
\usepackage{hyperref}
\hypersetup{colorlinks,citecolor=Blue,linkcolor=Red,urlcolor=Blue}
\usepackage{amsmath}
\usepackage{empheq}
\usepackage[usenames,dvipsnames]{color}

\allowdisplaybreaks 

\begin{document}
 
\title{The hot Jupiter period-mass distribution as a signature of in situ formation}  
\author{Elizabeth Bailey, Konstantin Batygin} 

\affil{Division of Geological and Planetary Sciences, California Institute of Technology, Pasadena, CA } 
\email{ebailey@gps.caltech.edu}

\newcommand{\Ham}{\mathcal{H}}
\newcommand{\G}{\mathcal{G}}
\newcommand{\appropto}{\mathrel{\vcenter{\offinterlineskip\halign{\hfil$##$\cr\propto\cr\noalign{\kern2pt}\sim\cr\noalign{\kern-2pt}}}}}
\newcommand{\Poincare}{{Poincar$\acute{\rm{e}}$}}

\begin{abstract}
More than two decades after the widespread detection of Jovian-class planets on short-period orbits around other stars, their dynamical origins remain imperfectly understood. In the traditional narrative, these highly irradiated giant planets, like Jupiter and Saturn, are envisioned to have formed at large stello-centric distances and to have subsequently undergone large-scale orbital decay. Conversely, more recent models propose that a large fraction of hot Jupiters could have formed via rapid gas accretion in their current orbital neighborhood. In this study, we examine the period-mass distribution of close-in giant planets, and demonstrate that the inner boundary of this population conforms to the expectations of the in-situ formation scenario. Specifically, we show that if conglomeration unfolds close to the disk's inner edge, the semi-major axis - mass relation of the emergent planets should follow a power law $a \propto M^{-2/7}$ \textemdash \,  a trend clearly reflected in the data. We further discuss corrections to this relationship due to tidal decay of planetary orbits. Although our findings do not discount orbital migration as an active physical process, they suggest that the characteristic range of orbital migration experienced by giant planets is limited.
\end{abstract} 

\maketitle
\section{Introduction} \label{sec1}
Speculation regarding the potential existence of giant planets that orbit their host stars in a matter of days dates back more than seven decades, to the proposed spectroscopic survey of \cite{Struve1952}. In retrospect, the remarkable lack of attention devoted to this possibility (in the 40 years that followed its publication, Struve's manuscript received 6 citations) can almost certainly be attributed to the stark contrast between the imagined nature of such objects and the expansive orbital architecture of our solar system. Accordingly, the 1995 discovery of the first hot Jupiter, 51 Pegasi b \citep{MayorQueloz95}, proved to be an immediate challenge to the hitherto conventional theory of giant planet formation \citep{Pollack96}, sparking considerable interest in reconciling the existence of Jupiter-like bodies on extremely close-in orbits with the theory of core-nucleated accretion. However, despite numerous efforts to conclusively resolve the problem of hot Jupiter formation, the origins of these remarkable objects remain imperfectly understood.

Generally speaking, the various formation pathways of Jovian-class planets at small orbital radii can be summarized into three broad categories: smooth migration, violent migration, and in-situ conglomeration. Within the framework of the first two scenarios, giant planet formation unfolds exclusively at large stello-centric distances (i.e. a few astronomical units) as originally imagined for the Solar System's giant planets \citep{BodenheimerPollack86}. Subsequently, upon conclusion of the primary accretion phase, the planet's orbital radius undergoes large-scale decay, shrinking by a factor of $\sim 10^2$ \citep{Linetal1996}. In the smooth migration picture, this is accomplished by dissipative interactions between the planet and its natal disk (via the so-called type-II mode of gas-driven migration; \citealt{KleyNelson12}), while the violent picture entails a sequence of events wherein the planet first attains a nearly parabolic trajectory (as a consequence of planet-planet scattering or the Lidov-Kozai mechanism; \citealt{BeaugeNesvorny12, Naoz11}) and then gets tidally captured onto a close-in circular orbit.

The in-situ model of hot Jupiter conglomeration \citep{KBAT} is markedly different from the picture described above in that the extent of orbital migration is assumed to be limited, and the vast majority of the planetary mass is imagined to accrete onto the planet locally (i.e. at a radial separation of order $\sim 0.1$ au or smaller). Importantly, in this case, core-nucleated instability is envisioned to be triggered by massive super-Earth type planets\footnote{The fact that the process of core-nucleated accretion is relatively insensitive to the temperature and pressure of the nebula, and can therefore proceed anywhere in the disk, was first demonstrated by the analytic calculations of \cite{Stevenson82}. More realistic numerical simulations of hot Jupiter conglomeration at $r\sim 0.05 \text{ au}$ are presented in \cite{KBAT}.}, which are strictly disallowed within the context of the traditional Minimum Mass Solar Nebula \citep{W, Hayashi81} but are found in great abundance around Sun-like stars by photometric and spectroscopic surveys \citep{Howardetal10,Mayoretal11,Batalhaetal13,DC13,DC15,F13,P13,M15,WF15}. We note, however, that for the purpose of our study, we remain completely agnostic as to the origins of the high-metallicity cores themselves: whether they too form locally \citep{ChiangLaughlin13, LeeChiang16,Boleyetal16} or instead get delivered to short-period orbits by (type-I) planet-disk interactions \citep{FoggNelson07, Bitschetal15} matters very little for the results that will follow.

In light of the relatively low occurrence rate of hot Jupiters ($\sim1\%$ for Sun-like stars; \citealt{Howardetal10}, \citealt{Gouldetal06}, \citealt{Wrightetal12}), it is not straightforward to determine which of the three aforementioned scenarios plays the dominant role in hot Jupiter generation. While observational signatures associated with each pathway have been widely discussed in the literature (see e.g. \citealt{WF15} for a review), these predictions typically entail some level of degeneracy. To this end, \cite{KBAT} have shown that the in-situ model is characterized by a key observational consequence - namely, that close-in Jovian planets should frequently be accompanied by (co-transiting as well as strongly inclined) super-Earth type companions. While circumstantial evidence has emerged for the existence of such companions \citep{Beckeretal15, Huangetal16}, \cite{SpaldingBat17} point out that the coexistence of hot Jupiters and low-mass planets is not strictly ruled out within the framework of the smooth migration paradigm, preventing a definitive distinction between the models. Furthermore, even spin-orbit misalignments, which were long touted as a marker of violent evolutionary histories \citep{Fab07, Naoz11}, have failed to conclusively inform the nature of hot Jupiter dynamical evolution, as numerous studies have shown that arbitrary stellar obliquities can naturally arise as a results of gravitational and magnetohydrodynamic disk-star interactions \citep{Lai1999, Bateetal10, SpaldingBat14,SpaldingBat15}.

With an eye towards resolving the ambiguity among the three categories of hot Jupiter formation models, here we examine the relationship between the masses of close-in giant planets and the distribution of their orbital periods. In particular, we argue that the observations signal a strong consistency with the in-situ formation scenario, suggesting that the extent of orbital migration suffered by this population of planets is unlikely to be particularly large. The remainder of the paper is structured as follows. In Section \ref{sec2}, we show that the inner boundary of the period-mass distribution of locally forming hot Jupiters is expected to follow a well-defined power law, and demonstrate empirical agreement between this relation and the observations. From there, we proceed to discuss tidal evolution. We present our conclusions in Section \ref{sec4}.

\section{Period-mass relation}\label{sec2}
The planetary mass as a function of the semimajor axis of the current observational census of extrasolar planets shown in Figure (\ref{f1}). Objects with confirmed (minimum) masses discovered via the radial velocity technique and transit observations are shown with blue and red points respectively. Transiting planets without direct mass measurements are shown as grey dots, and their masses are estimated using the mass-radius relationship of \cite{ChenKipping17}.

\subsection{The inner boundary}
\begin{figure*}
\centering
\includegraphics[width=0.98\textwidth]{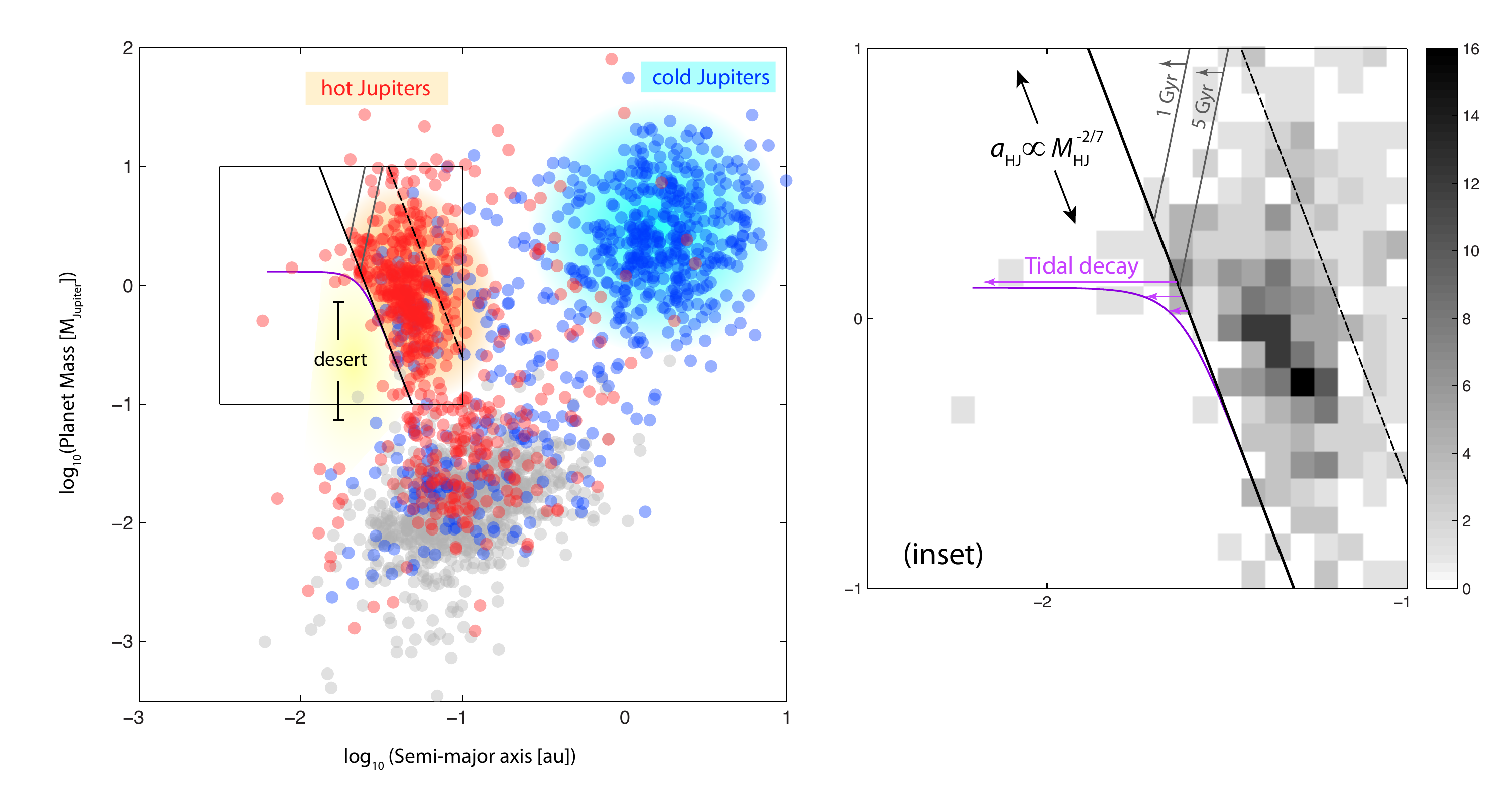}
\caption{The $a\propto M^{-2/7}$ relation derived for in-situ hot Jupiter formation shows empirical agreement with the lower boundary of the observed giant planet population in the $a-M$ diagram. \textbf{Left panel:} The cold Jupiter (blue shading) and hot Jupiter (red shading) populations are shown in relation to the giant planet ``desert'' (yellow shading). \textit{Blue points:} planets detected via the radial velocity technique, for which $M\sin i$ is plotted in lieu of $M$. \textit{Red points:} transiting planets with directly determined masses. Transiting planets with masses inferred from a mass-radius relation are shown as grey points. \textbf{Right panel (inset):} Density histogram in the $\log a-\log M$ plane. Maintaining the assumed T-Tauri star's surface field at $B\sim1\,$kG and varying the radius within the observed range yields lines that traverse the hot Jupiter population. Lines corresponding to $R_{\star} \sim 1.2R_{\sun}$ (solid) and $2R_{\sun}$ (dashed) bound the approximate lower and upper edges of the most populated region, respectively. Divergence from this empirical best fit line at short orbital radii agree with the tidal decay curve (purple) showing the evolution from the best fit line expected after $5$ Gyr of evolution. The grey lines illustrate the tidal decay isochrons described in the text.} 
\label{f1}
\end{figure*}

The inner edge of the $a-M$ diagram shown in Figure (\ref{f1}) has a rather well-defined profile, exhibiting a clear dependence on the planetary mass. Specifically, for planets less massive than $\sim 0.1M_{J}$, the boundary has positive slope, while the converse is true for more massive planets \citep{Mazeh16}. Given the four orders of magnitude spanned by the range of Figure (\ref{f1}), it is entirely plausible that the two dividing lines are carved by unrelated physical processes. 

The distribution of sub-Jovian ($M< 0.1M_J$) planets is almost certainly sculpted by photoevaporation \citep{OwenWu13, LopezFortney14}. Recasting the period-mass diagram into an irradiation-radius diagram, \cite{Lundkvistetal2016} have argued that the region of parameter space that exhibits a strong paucity of planets (the so-called sub-Jovian desert) is fully consistent with the effects of atmospheric mass loss. Moreover, the recent determination that the radius distribution of sub-Jovian planets is strikingly bimodal (as predicted by the photo-evaporation models; \citealt{Fultonetal17}) adds further credence to the notion that the origin of the positively sloped boundary in Figure (\ref{f1}) is rooted in radiative stripping of planetary envelopes.

Intriguingly, the same process cannot be invoked to explain the orbital architecture of hot Jupiters as a population. Models of atmospheric mass loss from highly irradiated giant planets \citep{Murray-ClayChiang09, Adams11} suggest that over the main-sequence lifetimes of their host stars, typical hot Jupiters will only lose $\sim 1\%$ of their total mass, altering the period-mass distribution to a negligible degree. As a consequence, a separate mechanism is needed to establish the negatively sloped boundary in Figure (\ref{f1}). Let us now examine the possibility that the observed distribution is nothing other than a relic of giant planet conglomeration at short orbital periods.

\subsection{In situ formation of hot Juipiters}

By now, it is generally accepted that the vast majority of hot Jupiters have formed via the core accretion pathway \citep{MillerFortney11}. Nevertheless, there is considerable uncertainty regarding the specific value of the critical core mass required to trigger runaway gas accretion at orbital radii smaller than $\sim 0.1$ au. In particular, 1D calculations of \cite{Ikoma, LeeChiang15, LeeChiang16} yield $2-3$ and $2-8$ Earth masses respectively, while simulations of \cite{Bod2000, KBAT} suggest a value closer to $15 M_{\Earth}$. Adding further uncertainty to this estimate, 3D hydrodynamic models of \cite{LambrechtsLega17} draw attention to the importance of global circulation within the Hill sphere for the determination of the energetics of this problem. 

The results of our study are largely insensitive to the specific characteristics of the high-metallicity core, as here we focus on the runaway accretion phase itself, during which the planet acquires most of its mass. Correspondingly, as a first step, it is worthwhile to consider the material budget of the inner disk. The amount of gas contained within $\xi = 0.1$ au of a \cite{Mestel1961}-type protoplanetary nebula with surface density profile $\Sigma = \Sigma_{0} (r_{0}/r)$ and $\Sigma_{0}=2000\text{ g cm}^{-2}$ at $r_{0}=1$ au is

\begin{equation}
\oint \int_{r_{\text{in}}}^{\xi} \Sigma r dr d\phi < 2\pi \Sigma_{0} r_{0} \xi \ll M_{J},
\end{equation}
\noindent where $r_{\text{in}}$ denotes the inner edge of the disk.

This simple estimate alone is sufficient to conclude that upon entering the runaway accretion regime, a locally forming hot Jupiter does not attain its final mass on a comparatively short (e.g. $\sim 10^4$ year) timescale. Instead, the gas must be delivered to the growing proto-planet by viscous accretion. Therefore, it is sensible to crudely express the hot Jupiter mass as
\begin{equation}\label{massprop}
M_{\text{HJ}} \sim \tau \dot{M} 
\end{equation}

\noindent where $\dot{M} \sim 10^{-8} M_{\sun}\text{ yr}^{-1}$ is the gas accretion rate at the inner edge of the disk \citep{Hartmannetal1998}, and $\tau \sim 10^{5} \text{ yr}$ is a characteristic accretion timescale (generally, some fraction of the disk lifetime). For the purposes of our rudimentary model, any dependence of the planetary accretion efficiency on mass simply translates into uncertainty of the free parameter $\tau$.

Within the framework of the in-situ model of hot Jupiter conglomeration, the smallest orbital radius where gas accretion can unfold is, roughly, the magnetospheric truncation radius of the disk. Importantly, like $M_{\text{HJ}}$, the truncation radius is also determined by $\dot{M}$. The expression for this length scale is well-known and is written as \citep{GhoshLamb1979a,Koenigl1991,Shuetal1994}
\begin{equation}\label{magrad}
 R_{t} \sim \Bigg( \frac{\mathcal{M}^2}{\dot{M}\sqrt{G M_{\star}} } \Bigg)^{2/7} 
\end{equation}
where $\mathcal{M}$ is the stellar magnetic moment,  $GM_\star$ is the star's standard gravitational parameter, and $\dot{M}$ is the disk accretion rate. Physically, $R_t$ is a characteristic radius at which viscous spreading of disk material is balanced by stellar magnetospheric torque acting upon the gas. 
 
Combining equations (\ref{massprop}) and (\ref{magrad}), we obtain the relation\footnote{Serendipitously, \cite{JWiz1980}, describing the onset of resonance overlap in the planar circular restricted three-body problem, \textit{also} derives a $-2/7$ power law $s_{\text{overlap}}\simeq\mu^{-2/7}$, where $s\simeq \sqrt{2/(3\Delta a)}$ and $\Delta a$ is the approximate separation of resonances. However, the underlying physics in these two cases is unrelated.}

\begin{equation}
a \sim \Bigg(  \frac{\mathcal{M}^{2} \tau}{M_{\text{HJ}}\sqrt{G M_{\star}}}\Bigg)^{2/7} \propto M_{\text{HJ}}^{-2/7}.\label{eqn:eq1}
\end{equation}

\noindent In $\log(a)-\log(M)$ space, this power law relation manifests as a line with slope $-2/7$. Figure (\ref{f1}) shows a line corresponding to the example T-Tauri parameters $M_{\star}\sim 1 M_{\sun}$ and $\mathcal{M} \equiv B_{\star}R_{\star}^{3}$ ($B_{\star} \sim 1 \text{ kG}$, $R_{\star} \sim 1.2 R_{\sun}$), in excellent agreement with the lower boundary of the hot Jupiter population. Intriguingly, keeping the other parameters constant while increasing the radius to $R_{\star}\sim2R_{\sun}$ yields an additional line, which, together with the aforementioned lower bound, envelops the approximate region of the parameter space most densely populated with observed hot Jupiters.

\subsection{Tidal evolution}

At greater masses and shorter periods, the observations appear to diverge from the $a \propto M^{-2/7}$ trend. As a resolution to this apparent disparity, let us consider the role of tidal evolution in shaping the hot Jupiter population. In particular, we follow the formalism outlined in \cite{MD99} for the standard case of a planet moving on a circular, equatorial orbit with mean motion $n$, around a star rotating with angular speed $\Omega$. For the case $\Omega<n$, the tidal bulge induced on the star by the planet lags behind the planet's orbit, leading to orbital energy loss and consequent decay of the semimajor axis. 

The contraction of hot Jupiter semimajor axes is predicted by the well-established formula \citep{MD99}:

\begin{equation}
\frac{da}{dt} = -\frac{3 k_{2\star}}{Q_{\star}}  \frac{M_{\text{HJ}}}{M_{\star}}  \Big( \frac{C_{\star}}{a}\Big)^{5} an.
\end{equation}

\noindent where $k_{2\star}$ is the tidal Love number of the star (equal to 0.01 for an $n=3$ polytrope \citep{BatAdams13}, appropriate for a fully radiative body, and $Q_{\star}$ is the stellar quality factor, typically estimated to be roughly $\sim 10^{5}\text{ to }10^{6}$ \citep{Levrardetal09}. The stellar mass is represented as $M_{\star}$, and the stellar radius is denoted as $C_{\star}$. Rearrangement of this equation and integration with respect to $a$ and $t$ yields an equation for the final semimajor axis $a_{f}$ in terms of initial semimajor axis $a_{i}$ and total evolution time $t$:

\begin{equation}\label{Eq2}
a_{f}= \Bigg( a_{i}^{13/2} - t\frac{13}{2} \frac{3 k_{2\star}}{Q_{\star}}  \frac{M_{\text{HJ}}}{M_{\star}}  C_{\star}^{5} \sqrt{GM_{\star}}\Bigg)^{2/13}.
\end{equation}

Imagining orbital decay to unfold over a typical system lifetime of $\sim5$ Gyr, originating from initial values of $(a, M)$ defined by the best-fit line found at the boundary of the hot Jupiter population, we obtain a tidally corrected inner boundary, which is shown in Figure (\ref{f1}) as a purple curve. Remarkably, tidal evolution appears to fully explain the bulk of trend-crossing hot Jupiters. While the expected number of these tidally decaying planets is difficult to predict precisely, due to observational biases inherent to the current observational sample of hot Jupiters and the sensitive dependence of tidal evolution on the initial conditions, forthcoming results of the \textit{Transiting Exoplanet Survey Satellite (TESS)} mission are expected to bring this population of in-falling planets into sharper focus.

Finally, we consider the role of tides in shaping the hot Jupiter population at even greater masses. Specifically, by rearranging (\ref{Eq2}), we obtain, for a given time span, an expression for the initial radius from which a planet of given mass decays to the Roche limit. Examples of the resulting curves, for $1$ and $5$ Gyr, are shown in grey (Figure \ref{f1}). Notably, these tidal decay isochrons agree with the approximate boundary of the hot Jupiter population for $M\gtrsim M_{J}$.

\section{Conclusion}\label{sec4}
At the dawn of exoplanetary observations, the conceptual foundation of planet formation was built upon the lone case of the Solar System. Given the paltry mass of the terrestrial planets and the lack of material orbiting interior to Mercury, it was thought that planet formation was generally inactive at short orbital radii \citep{Cameron1988,Rafikov06}. To alleviate the ostensible paradox brought about by the discovery of hot Jupiters, migration mechanisms were invoked to explain how giant planets could be delivered inward from distant, Jupiter-type orbits \citep{Linetal1996}.

Today, the landscape of exoplanet detections foretells a very different story \citep{LaughlinLissauer15}. It is observationally well established that a generic outcome of the planet formation process is short-period super-Earths, the most massive of which can successfully trigger rapid gas accretion and become gas giants, if allowed to reside within their natal nebulae for $\sim1$ Myr \citep{Bod2000, KBAT}. In fact, given the remarkable scarcity of close-in gas giants relative to sub-Jovian short-period planets, all that is needed to reproduce the vast majority of the hot Jupiter population in situ is for $\sim1\%$ of young super-Earths to enter the runaway regime of conglomeration before dissipation of their protoplanetary nebulae.

In this work, we have explored the in-situ formation scenario of hot Jupiters further, and demonstrated that a bounding relation $a \propto M_{\text{HJ}}^{-2/7}$ is expected to manifest if a significant fraction of these objects formed locally. Intriguingly, we find that the slope of this power law is in excellent empirical agreement with the lower edge of the hot Jupiter population on the $a-M$ diagram, with corrections from tidal dissipation playing a secondary role (Figure \ref{f1}). Accordingly, this finding yields further support to the hypothesis that in-situ formation accounts for a considerable fraction of hot Jupiters.

We note that, in addition to typical short-period Jovian planets that reside on nearly circular orbits, there exist numerous instances of highly eccentric hot Jupiters with exterior companions, for which the most simple explanation is that they are undergoing the final circularization phase of violent (possibly Lidov-Kozai) migration \citep{WM, Fab07}. While these objects certainly do not fit into the picture presented herein, \cite{Dawson} have demonstrated that only a minority of hot Jupiters could have formed via this high-eccentricity pathway, weakening the case for this flavor of orbital transport as a dominant route for hot Jupiter production (see also \cite{Ngoetal16}). Moreover, unlike the upper boundary of the hot super-Earths in the $a-M$ diagram (which is adequately explained as resulting from photoevaporation; \citealt{OwenWu13}, \citealt{LopezFortney14}), the mass-period relationship governing the sharp lower boundary of the hot Jupiters has so far evaded migratory explanations \citep{OwenLai18}.

Despite the aforementioned correspondence between the in-situ $a-M$ relation (\ref{eqn:eq1}) and the data, it is clear that treating migration as utterly non-existent in planet formation theory is as extreme as demanding that migration must necessarily be long-range. To the contrary, there is no doubt that, at least to some extent, giant planet migration plays a role in shaping planetary systems. For example, mean motion resonances found in systems such as GJ 876 \citep{Marcyetal01} are almost certainly a product of convergent migration \citep{LeePeale02}. Moreover, our own solar system holds distinct markers of past giant planet migration, not least of all being the notion that the terrestrial planets are best reproduced in models that include successive inward and outward migration of Jupiter over several au (the so-called ``Grand Tack;'' \citealt{Walshetal11}, see also \citealt{BatyginLaughlin15}). Importantly, however, systems that show evidence of migratory sculpting typically require only short-range orbital transport. Thus, our results cumulatively suggest that long-range migration of giant planets is likely to be the exception rather than the rule.

\section{Acknowledgments} We wish to thank Dan Fabrycky, Greg Laughlin, Eugene Chiang, and Matt Holman for useful conversations, and the anonymous reviewer for useful comments. This research has utilized the NASA Exoplanet Archive, which is operated by the California Institute of Technology, under contract with the National Aeronautics and Space Administration under the Exoplanet Exploration Program.

\end{document}